\documentclass[12pt,preprint]{aastex}

\begin{document}


\title{Discovery of New Ultracool White 
Dwarfs in the Sloan Digital Sky Survey}


\author{Evalyn Gates\altaffilmark{1}, Geza Gyuk\altaffilmark{1,2}, 
Hugh C. Harris\altaffilmark{3}, Mark Subbarao\altaffilmark{1,2},
Scott Anderson\altaffilmark{4},\\ 
S. J. Kleinman\altaffilmark{5}, James Liebert\altaffilmark{6}, 
Howard Brewington\altaffilmark{5}, J. Brinkmann\altaffilmark{5}, 
Michael Harvanek\altaffilmark{5}, 
Jurek Krzesinski\altaffilmark{5,7}, Don Q. Lamb\altaffilmark{1},
Dan Long\altaffilmark{5}, Eric H. Neilsen, Jr.\altaffilmark{8}, 
Peter R. Newman\altaffilmark{5}, Atsuko Nitta\altaffilmark{5}, and
Stephanie A. Snedden\altaffilmark{5}}

\altaffiltext{1}{The Univ. of Chicago, Department of 
Astronomy \& Astrophysics, 5640 S. Ellis Ave., Chicago, IL 60637}
\altaffiltext{2}{Adler Planetarium \& Astronomy Museum, 
1300 S. Lake Shore Drive,
Chicago, IL 60605}
\altaffiltext{3}{U.S. Naval Observatory, Flagstaff Station, 
P.O. Box 1149, Flagstaff, AZ 86002-1149}
\altaffiltext{4}{Department of Astronomy, University of Washington,
Box 351580, Seattle, WA 98195}
\altaffiltext{5}{Apache Point  Observatory, P.O. Box
59, Sunspot, NM 88349-0059}
\altaffiltext{6}{Steward Observatory, Univ. of Arizona,
933 N. Cherry Ave., Tucson, AZ 85721}
\altaffiltext{7}{Mt. Suhora Observatory, Cracow Pedagogical University,
ul. Podchorazych 2, 30-084 Cracow, Poland}
\altaffiltext{8}{Fermi National Accelerator Laboratory, P.O. Box 500,
Batavia, IL 60510}

\begin{abstract}
We report the discovery of five very cool white dwarfs in the Sloan Digital
Sky Survey (SDSS).  Four are ultracool, exhibiting strong collision
induced absorption (CIA) from molecular hydrogen and are similar in color
to the three previously known coolest white dwarfs, SDSS J1337+00, LHS
3250 and LHS 1402. The fifth, an ultracool white dwarf candidate, shows
milder CIA flux suppression and has a color and spectral shape similar to
WD 0346+246.  All five new white dwarfs are faint ($g > 18.9$) and 
have significant proper motions.
One of the new ultracool white dwarfs, SDSS J0947, appears to be in a 
binary system with a slightly warmer 
($T_{eff} \sim 5000K$) white dwarf companion.   
\end{abstract}




\section{Introduction}

White dwarf stars with $T_{eff} < 4000 K$ are of great interest for
several reasons.  End-stage remnants of main sequence stars with masses
less than about $8 M_{\odot}$, they represent some of the oldest objects
in the galaxy. As such they give direct information about star formation
during the Galaxy's earliest epochs.  Since white dwarfs continue to cool
and fade with time, the very coolest can be used place lower limits on the
ages of various galactic components. In addition, recent microlensing
searches have suggested that there may be a significant population of
white dwarfs in the galactic halo \citep{alc00}. These fossil remains of 
an ancient
stellar population could offer a window into the early stages of the
galaxy and its formation.

Recent theoretical progress in understanding the evolution of white 
dwarfs \citep{han98,sau99} at temperatures below $6000 K$ has helped to refocus
the search for these objects. Ultracool white dwarfs ($T_{eff} < 4000 K$)
with hydrogen in the atmosphere will exhibit a unique spectral signature due to
collision induced absorption (CIA) by molecular hydrogen.  Such absorption
results in a bluer spectrum, with a significant flux suppression red-ward
of about 6000 \AA, relative to a blackbody SED.

To date, only three ultracool white dwarfs with strong CIA flux
suppression have been observed -- LHS 3250 \citep{har99,har01,opp01b}, 
SDSS J133739.40+0001428 
(hereafter referred to as
SDSS J1337; Harris et al. 2001) and LHS 1402 \citep{opp01,sal03}. 
These studies have dramatically
confirmed the general predictions of the models. However, detailed
agreement between model and observed spectra is very poor, making accurate 
temperature and age estimates impossible. For example, model predictions
vary for different atmospheric composition and mass, but all predict a
spectral feature due to $H_2$ at about 7500 - 8000 \AA.  None of the
observed spectra have shown any evidence for this feature.

There is a second group of ultracool white dwarfs that exhibit milder CIA
flux suppression. This group includes WD 0346+246 
\citep{ham97,hod00,opp01b}, LHS 1126
\citep{ber94,ber97}, and GD392B \citep{far04},
all of which have data at wavelengths above 10,000 \AA.  
These studies found a significant 
flux deficiency in the near-infrared (1-2 microns),
which confirms the presence of CIA in these stars and thus
their classification as ultracool white dwarfs.
There are also a handful of cool white dwarfs which may belong to this
second group, including a wide binary pair of white dwarfs, SSSPM
J2213-7514 and SSSPM J2213-7515 \citep{sch02}, F351-50 \citep{iba00} 
and CE 51 \citep{rui01},
which is in a binary system with an M star.  
However, no infrared data exist for these
objects as yet, and their colors and optical spectra are not conclusive 
evidence that they are ultracool.

The unusual colors of ultracool white dwarfs led to predictions
\citep{har99,han00} that they should be detectable in the Sloan Digital
Sky Survey (SDSS; York et al. 2000, Abazajian et al. 2003,
Abazajian et al. 2004, Gunn et al. 1998, Stoughton et al. 2002).  
The colors of these coolest white dwarfs
lie in a region of color-color space that is distinct from that of most
stars including higher temperature white dwarfs.  This region is
sparsely populated and strongly overlaps the region where high
redshift ($z>3$) QSOs are found \citep{ric02}.  
In fact, two ultracool white dwarfs were
picked up in the commissioning data of the SDSS, one of which was a new
discovery (SDSS J1337+00) and one which was previously known
(LHS3250) \citep{har01,luy76}.  Both of these objects were targeted as QSOs.
However, cool white dwarfs with pure helium atmospheres or
those with weak CIA are less likely to be targeted for spectral
observation since these will have colors which 
overlap the locus of ordinary stars.
 
In this paper we report the discovery of new ultracool white dwarfs in
the SDSS.  We have found four stars exhibiting strong CIA, similar to LHS 3250
and SDSS J1337+00, more than doubling the number of known ultracool 
white dwarfs
with strong flux suppression.  We also report the discovery of a fifth
star, an ultracool white dwarf candidate, which shows a milder suppression and
has colors closer to those of the second group exemplified by WD 0346+246.

\section{Observations}

Ultracool white dwarfs exhibiting strong CIA 
have a high probability of being selected for
spectroscopic observations by SDSS as possible high redshift quasars 
as well as cool white dwarfs \citep{har01}.
We have performed a thorough search of
all SDSS \citep{pie03,smi02,hog01} 
spectral data available as of April 2004 (approximately 
$50\%$ of the total number of
spectra that will ultimately be targeted by SDSS). All spectra were
obtained with the SDSS 2.5 m telescope multifiber spectrographs,
which cover 3800 - 9200 \AA, at a spectral resolution of 1800
\citep{yor00}.
Because the spectra of
ultracool white dwarfs are featureless they are classified as unknown by
the SDSS spectroscopic pipeline, and we visually examined all SDSS unknown
spectra.  The unique spectral shape of the ultracool white dwarfs
stood out dramatically from the other objects in this category. 

The five new ultracool white dwarfs reported here are 
SDSS J094722.98+445948.5, SDSS J122048.65+091412.1, 
SDSS J100103.42+390340.4, SDSS J140324.66+453332.6 and 
SDSS J085443.33+350352.7 (hereafter referred to as SDSS J0947,
SDSS J1220, SDSS J1001, SDSS J1403 and SDSS J0854, respectively).
Finding charts for each new star are
shown in figure 1.  Their positions, proper motions and
colors are given in Table 1.
We also recovered the ultracool
white dwarfs previously found (or recovered) in SDSS, LHS 3250 and SDSS
J1337.

Preliminary proper motions were calculated using the SDSS and USNO-B
catalog positions, but in at least two cases these proper
motions were clearly unreliable. We hence returned to the POSS I, II and
SDSS images and obtained proper motion measurements by direct
reduction. The five new objects all have highly significant proper
motions, and SDSS J0947 has a proper motion companion at a
distance of 20" to the northeast. The SDSS colors and luminosity of this
object suggest that it is also a white dwarf, though of a higher
temperature.

The colors of the new white dwarfs are shown in Figure 2.  For comparison
we have also plotted the other known ultracool white 
dwarfs\footnote{SDSS colors for ultracool white dwarfs without SDSS data
were estimated using the photometric transformations of \citet{fuk96},
expected to be accurate to about 0.1 mag. The one exception to this was
the $r - i$ color of LHS 1402 which we extracted from its 
shape-calibrated spectrum
(Oppenheimer, private communication 2004). We were unable to
obtain sufficient color data for F351-50 to estimate SDSS colors.} 
as well as the
locus of points for a sample of normal white dwarfs in the SDSS 
\citep{kle04}.    The unusual colors of the four new
stars with strong CIA flux suppression - SDSS J0947, SDSS J1220, SDSS
J1001 and SDSS J1403 -- are evident in this plot.  
Based on their colors, which fall well
apart from the locus of normal white dwarfs, two of these objects
were targeted for spectral observation as possible QSO candidates and 
two were targeted by the category SERENDIPITY-DISTANT as
having colors distant from the stellar locus.  
SDSS J0854, which exhibits milder CIA suppression, lies much closer to the
locus of warmer white dwarf colors and was targeted as a carbon star.
However, the spectra for all five new stars are distinctive as can be seen
in Figure 3.  All are featureless and show a noticeable flux suppression
at wavelengths longer than about 6000 \AA.  This suppression is
especially strong in SDSS J0947, SDSS J1001, SDSS J1220 and SDSS J1403.  
The spectra
of SDSS J0947 and SDSS J1001 are very similar to each other and also to
LHS 3250.  SDSS J1403 exhibits the most severe flux suppression, 
while the spectrum of SDSS J1220 is somewhat more sharply peaked than
the others, with a steeper slope blueward of the peak.

\section{Temperature and Atmospheric Composition}

Model atmosphere calculations indicate that strong CIA requires a
temperature below 4000 K, and previous studies of these stars have favored
atmospheres dominated by helium, although 
even helium-rich models fail to accurately reproduce 
the observed spectra in detail (Bergeron \& Leggett 2002). 
The small number of ultracool white dwarfs that have been observed
has also hampered progress toward a better understanding of their
composition and properties. However, the addition of five new 
new stars allows
us to begin a rough classification of these objects based on their
colors and amount of CIA suppression.
There are two rough groupings of ultracool white dwarfs
discovered so far. (However we note that this grouping may
not necessarily indicate any underlying physical distinction
between the stars other than temperature.)

In the first group are SDSS J1337, LHS 3250, SDSS J0947,
SDSS J1220, SDSS J1001, LHS 1402 and SDSS J1403.
All seven lie in the same region of color-color space, well below the
locus of normal white dwarfs,
and exhibit strong CIA suppression.  This indicates temperatures
below about $4000 K$. Previous temperature estimates for
LHS 3250, SDSS J1337 and LHS 1402 have been in the broad range of
$2000 - 4000 K$,  and our
four new white dwarfs are also likely to have temperatures in this
range. SDSS J1001 is 
likely to be similar in temperature to SDSS J1337, while SDSS J0947
is probably a bit warmer. If the relative position in color-color space
indicates a progression downward in temperature as the cool white dwarfs
fall farther from the cool end of the normal white dwarf locus, 
SDSS J1403 may be the coolest white dwarf yet discovered.
While the spectra
of SDSS J1001, SDSS J0947 and SDSS J1403 are similar to LHS 3250,
trigonometric parallaxes and detailed model fitting will be necessary to
determine if they are also likely to be overluminous, He-rich, 
low mass binaries as suggested for LHS 3250 \citep{har99,ber02}.
 
The fourth new star, SDSS J1220, shows significant CIA suppression, but
lies a bit further apart in color space from the others in the first 
grouping.  Its spectral
shape exhibits a relatively sharp peak, with a steep
fall-off in flux both before and beyond about 6000 \AA, 
which may indicate a different
atmospheric composition from the others.  However, until more
detailed model comparisons can be made no strong conclusions
are possible.  We expect that this object also lies in the
broad temperature range of $2000 - 4000 K$, possibly toward the
cooler end.

The second grouping of ultracool white dwarfs includes
WD 0346+24, F351-50, GD392B and LHS 1126, along with the possible ultracool
white dwarf candidates SSSPM J2231-7514, SSSPM J2231-7515, and
the new star SDSS J0854.  All of these stars lie close to the normal white
dwarf locus in color space and cannot be distinguished from the 
locus of ordinary stars based on color alone. They exhibit milder CIA flux
suppression, and may have temperatures closer to $4000 K$ than those
in the first grouping.  

Like all three of the previously
known ultracool white dwarfs, none of our new
white dwarfs exhibits the $H_2$ feature at roughly 8000 \AA
(which is more pronounced in models with pure H atmospheres), as
predicted by the models 
despite the use of the latest opacities calculated for H$_2$
(Borysow, J{\o}rgensen, \& Fu 2001; J{\o}rgensen et al. 2000).
Some improvement in either the opacities or the models
(e.g. Kowalski \& Saumon 2004, Iglesias et al. 2002) 
will be needed to fit these spectra and extract more accurate estimates of
the temperatures and compositions of these stars.

\section{Disk or Halo?}

Assigning membership of these new stars to a particular component of the
galaxy is problematic.  Estimating distances, and thus tangential
velocities, is difficult with so few known objects of this type.  
No color
magnitude relation (CMR) yet exists and because of the dramatically
different colors of these objects, the CMR for normal cool white dwarfs is
inappropriate. Extracting absolute magnitude estimates from comparison to
theoretical models is unreliable. 
For example, LHS
3250 has a parallax distance measurement which implies
an absolute magnitudes much brighter than predicted by models of normal mass,
hydrogen atmosphere white dwarfs that have cooled to temperatures where
CIA becomes significant in the optical spectra.

The parallax distance measurement for LHS 3250 allows an accurate
determination of M$_V = 15.72 \pm 0.04$ \citep{har99}.  If we assume a
similar absolute magnitude for the new ultracool white dwarfs in the first
group, we find, for SDSS J0947, a distance of $d \sim 47$ pc and a
corresponding $v_{tan}\sim 20$ km s$^{-1}$.  
For SDSS J1001, we find $d \sim 64$ pc
and $v_{tan}\sim 107$ km s$^{-1}$, while for SDSS J1403
we obtain $d \sim 44$ pc
and $v_{tan}\sim 60$ km s$^{-1}$.  Likewise, SDSS J1220, which has the highest
proper motion, has $d \sim 64$ pc and $v_{tan}\sim 154$ km s$^{-1}$.
Furthermore, SDSS J0947 has a companion with common proper motion: SDSS
J094724.45 +450001.8 has colors consistent with a WD of
$T_{eff} \sim 5000 K$ at a distance of about 60 pc with a tangential
velocity of 25 km s$^{-1}$.  
If this distance is correct, then like LHS 3250 the
absolute magnitude of SDSS J0947 is brighter than that predicted by the
models for a cool halo WD of normal mass, and it adds to the evidence that
it is a disk star with a large radius and small mass.
  
We can use a similar approach for ultracool white dwarfs in the second
group, assuming the absolute magnitude of SDSS J0854 is similar to that of
WD 0346+246, which has $M_V = 16.8\pm 0.3$ based on its parallax distance
\citep{ham99}.  For SDSS J0854 we
find $d \sim 41$ pc and $v_{tan} \sim 43$ km s$^{-1}$.

A more conservative approach for all of the new ultracool white dwarfs is
to consider a value for the absolute magnitude of M$_V = 16.5\pm 1.0$,
following \citet{sal03}.  This range encompasses a wide suite of model
predictions for an $m= 0.6 M_{\odot}$, pure H or mixed H/He white dwarf
which exhibits CIA suppression.  The results are given in Table 1.
Based on these estimates of $v_{tan}$, it is clear that SDSS J0947 and
SDSS J0854 are members of the galactic disk, 
and SDSS J1403 probably is as well. 
SDSS J1001 may be
either disk or halo; however, if it has the bright M$_V <  16$ required for
the larger values of $v_{tan}$, it must also have a large radius and
low mass similar to that implied for LHS 3250 \citep{har01}.  
SDSS J1220 is likely to be a halo white dwarf, which
makes its unusual colors and steep flux suppression even more interesting.
It is the only ultracool white dwarf known that can be a halo star
with normal mass that has cooled to a temperature substantially lower
than WD 0346+246.
Ultimately, of course, trigonometric parallax 
measurements will be necessary to fully understand these stars. 

Based on the six stars with strong CIA detected in 
$\sim 4330$ deg$^2$ of sky observed for SDSS spectra through April 2004,
we find 0.0014 deg$^{-2}$ ultracool WDs with $i<20.2$
(the magnitude limit for selection of QSO candidates),
or approximately $R<19.8$.
This density is somewhat higher than found by Oppenheimer
et al. (2001a) who found one star (LHS 1402) in 4200 deg$^2$
with $R<19.8$ and $\mu > 330$ mas yr$^{-1}$
(three of our six stars are within these limits),
and it is consistent with the LHS Catalog that has two stars
with $R<18$ and $\mu > 500$ mas yr$^{-1}$ (although the LHS
limiting magnitude varies over the sky).
To estimate their space density, we note that nearly all
ultracool WDs will be selected for spectra by one or both
of the QSO selection algorithms.
The magnitude limits are $i<19.1$ for low-z QSO candidates
and $i<20.2$ for redder high-z candidates.
In fact, all six strong-CIA stars were flagged by the
QSO selection procedure, and four were assigned fibers as
QSO candidates\footnote{
Of the other two, SDSS J1001 was too faint for low-z selection
so was flagged as QSO\_MAG\_OUTLIER, and SDSS J1403 had colors
in the A-star reject box so was flagged as QSO\_REJECT.
Both were observed anyway by the backup SERENDIPITY\_DISTANT.
Because SERENDIPITY and other non-QSO selection
categories are not complete, we omit these two stars from
the calculation of space density.  The faint magnitude limits for 
these categories imply large discovery volumes and thus we expect 
the density contributions will be small in any case.}.
Assuming that these four stars are all like LHS 3250 with
M$_r = 15.47$, and summing the inverse of their potential discovery volumes
gives a space density of $3.0 \times 10^{-5}$ pc$^{-3}$.
This value is very uncertain because of the uncertain distance
and luminosity of most of the stars, but is similar to the
density of the disk white dwarf luminosity function at the faintest
measured luminosity bin (Leggett, Ruiz, \& Bergeron 1998). We note that
our estimate reflects a mix of various galactic components.

Finally, we note that the status of SDSS J0947 as a member of a binary
system is far from unique.  CE 51 has a main sequence companion, 
GD392B has a probable white dwarf companion and SSSPM
J2231-7514 and SSSPM J2213-7515 are a wide binary pair.  Thus, if these
latter three stars are confirmed as ultracool, $\sim 30\%$ of the known
ultracool white dwarfs would be in binary systems.

In summary, we have discovered five new ultracool white dwarfs in the SDSS.  
Four have colors and spectra indicating strong CIA, and one appears to have the
strongest CIA of any star discovered to date.
Only one star has a proper motion sufficiently large 
that it is likely to be a halo star with low luminosity and normal mass.
Three of the others have smaller proper motions indicating that they
are probably disk stars with younger ages, higher luminosities and 
smaller masses, and one of these three has a warmer white dwarf 
companion with a photometric distance supporting the small-mass 
interpretation. The fifth star has a proper motion that could be either 
disk or halo, but for higher tangential velocities it must also have
a high luminosity and low mass.
None of the spectra show bands of H$_2$ predicted by 
current white dwarf
atmosphere models, and we find a rough estimate of the density of ultracool 
white dwarfs of
about $3.0 \times 10^{-5}$ pc$^{-3}$.

\acknowledgments

The authors would like to thank Ben Oppenheimer for generously providing
the spectral data for LHS 1402.  G.G. and M.S. would like to thank the
Brinson Foundation for generous support.  This research has made use of
the USNOFS Image and Catalogue Archive operated by the United States Naval
Observatory, Flagstaff Station. Funding for the creation and distribution 
of the SDSS Archive has been provided by the Alfred P. Sloan Foundation, 
the Participating
    Institutions, the National Aeronautics and Space Administration, the
    National Science Foundation, the U.S. Department of Energy, the
    Japanese Monbukagakusho, and the Max Planck Society. The SDSS Web
    site is http://www.sdss.org/.
The SDSS is managed by the Astrophysical Research Consortium (ARC)
    for the Participating Institutions. The Participating Institutions
    are The University of Chicago, Fermilab, the Institute for Advanced
    Study, the Japan Participation Group, The Johns Hopkins University,
    Los Alamos National Laboratory, the Max-Planck-Institute for
    Astronomy (MPIA), the Max-Planck-Institute for Astrophysics (MPA),
    New Mexico State University, University of Pittsburgh, Princeton
    University, the United States Naval Observatory, and the University
    of Washington.

\clearpage



\begin{figure}
\plottwo{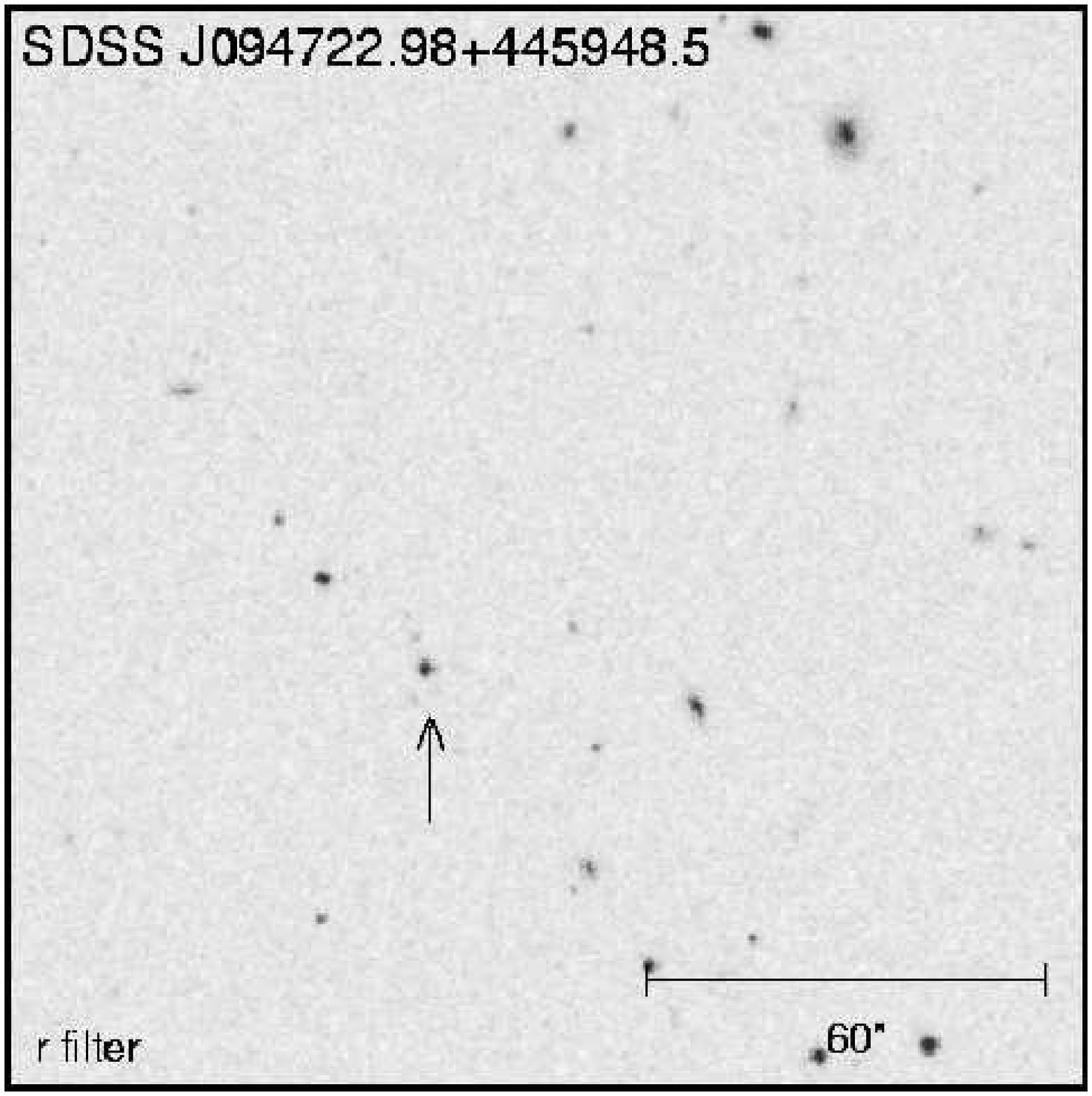}{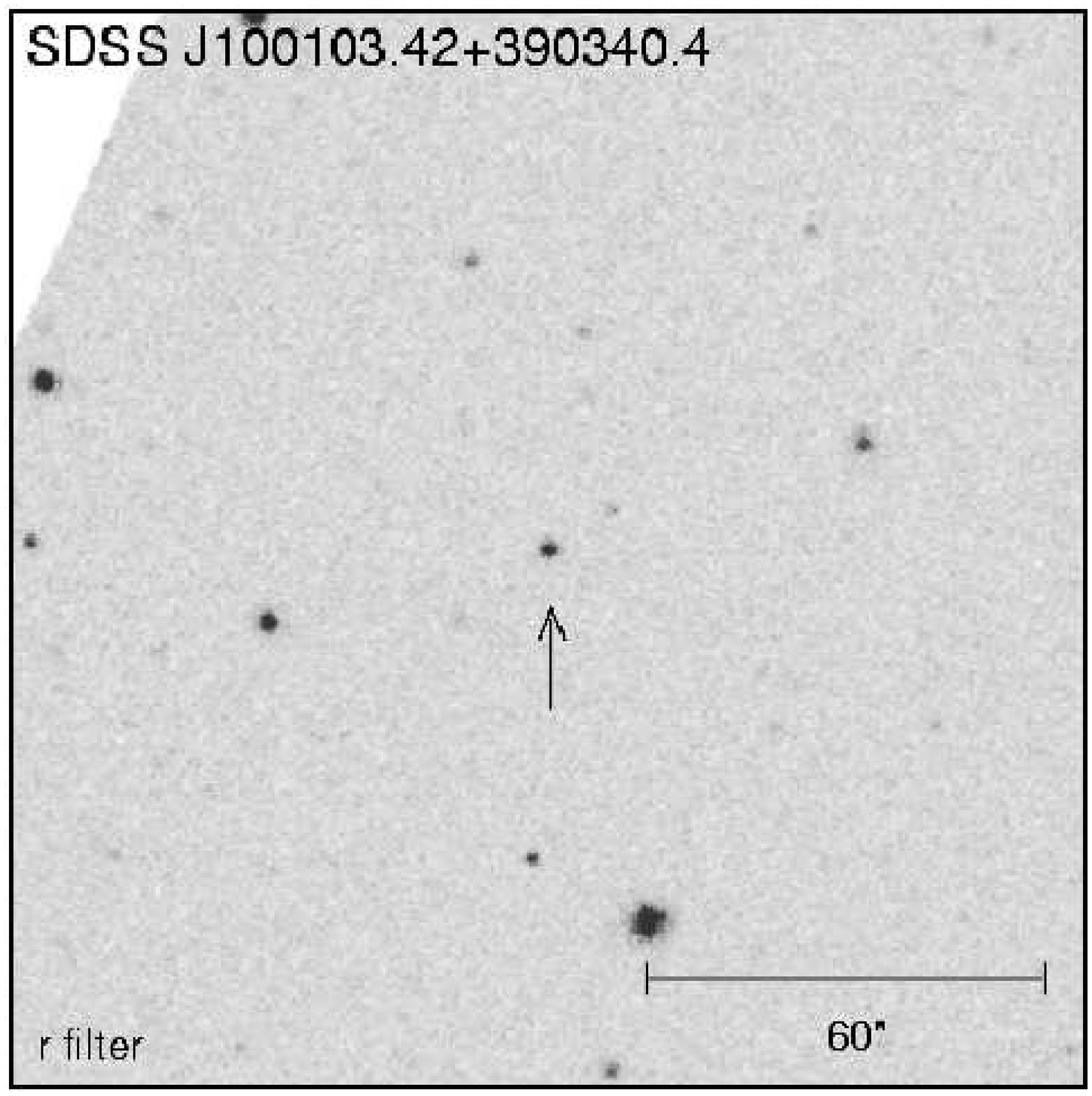}
\caption{(a) Finding charts for new ultracool white dwarfs 
from
r band SDSS images. North is up, east is to the left. Epochs are
given in Table 1.}
\end{figure}
\setcounter{figure}{0}
\begin{figure}
\plottwo{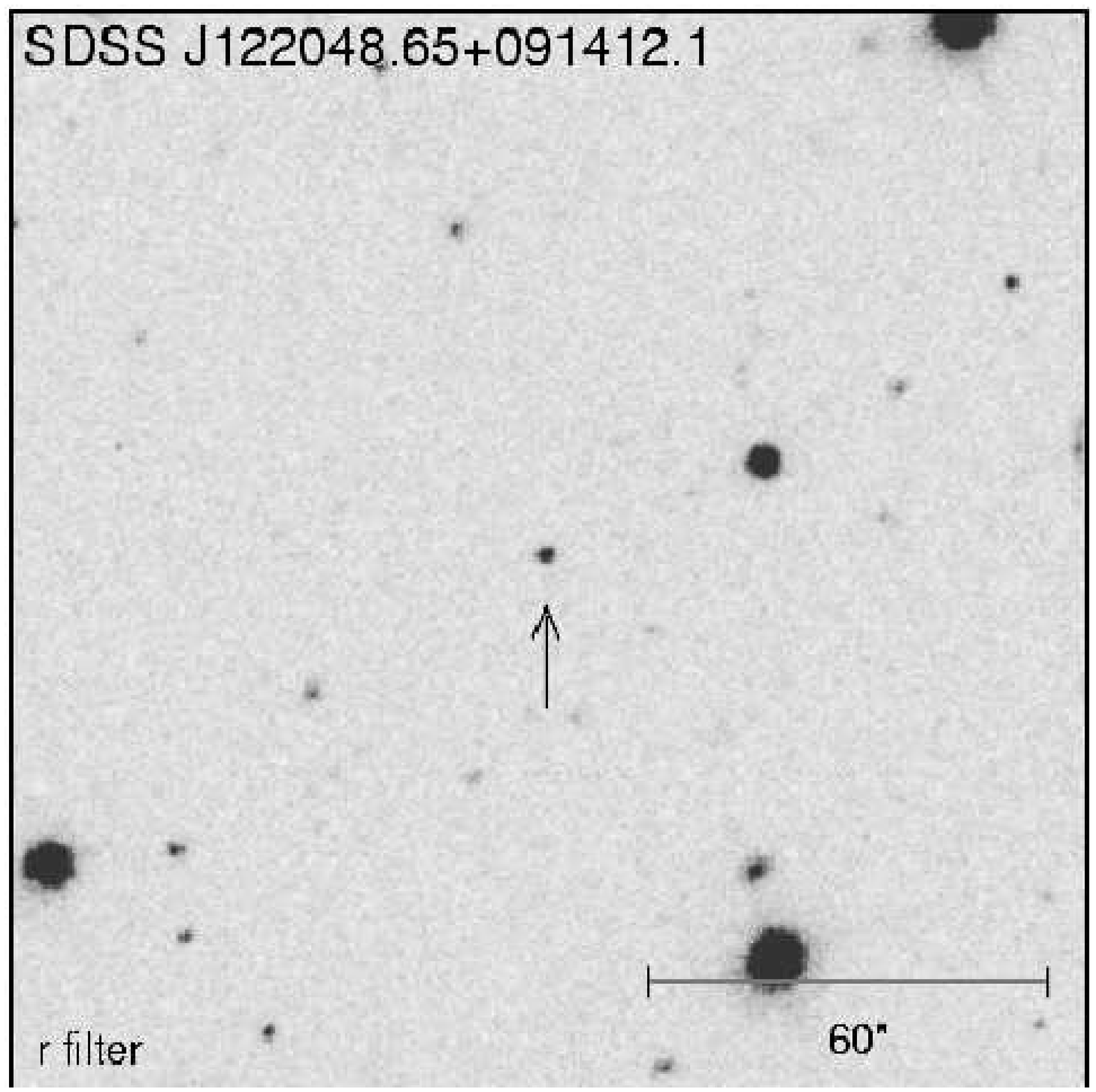}{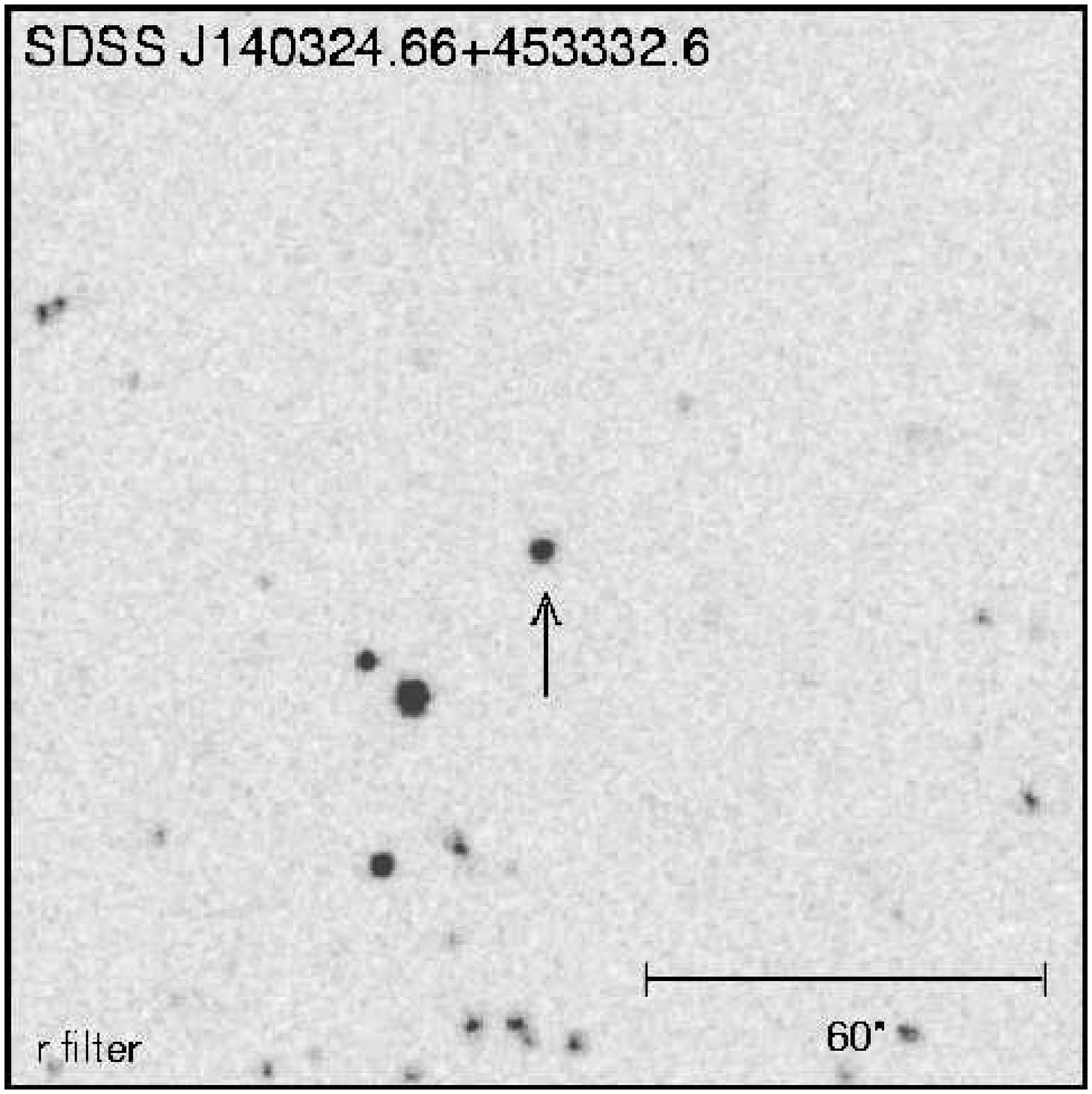}
\caption{(b)}
\end{figure}

\setcounter{figure}{0}

\begin{figure}\nonumber
\epsscale{0.5}
\plotone{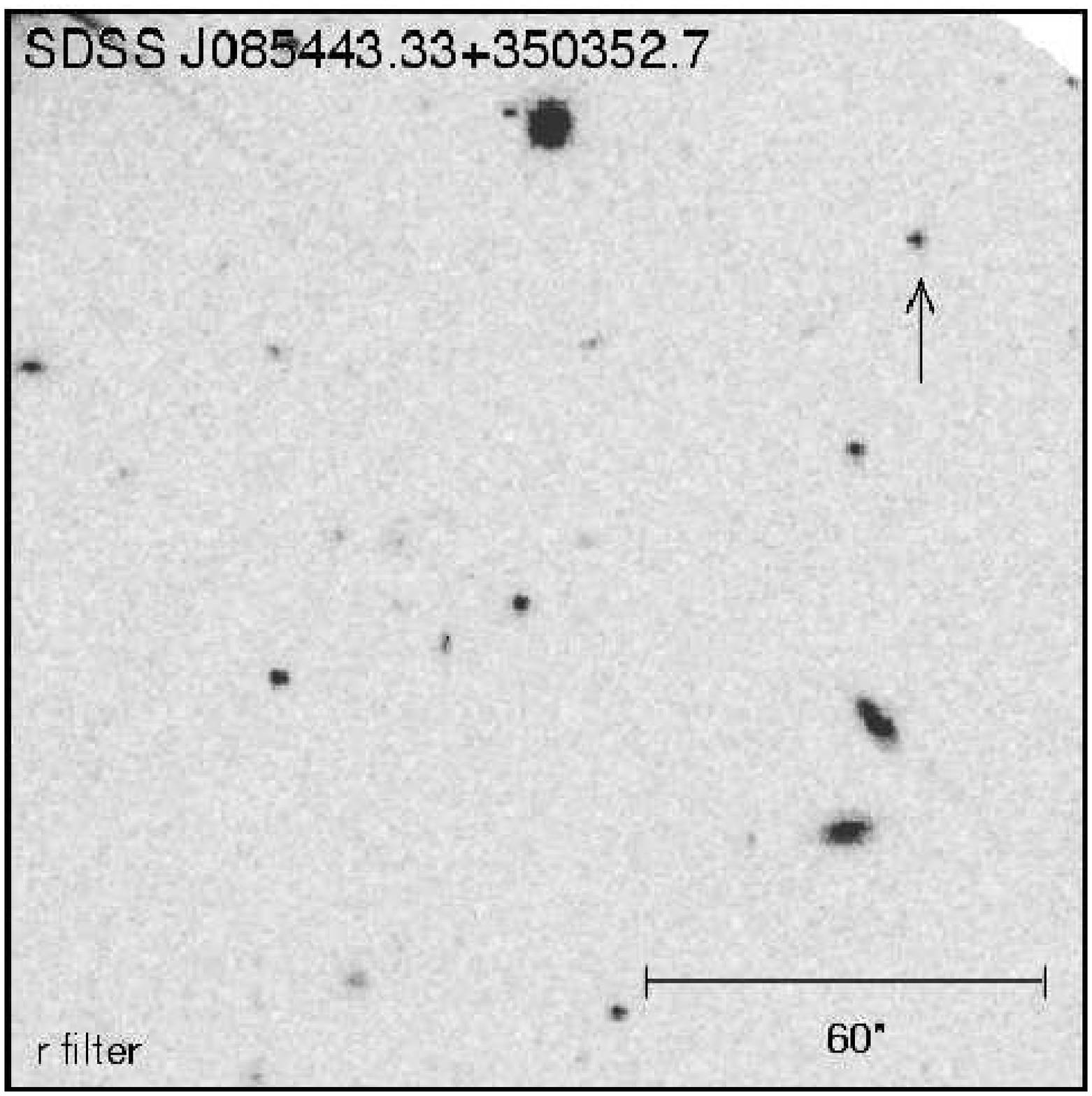}
\caption{(c)}
\end{figure}

\clearpage

\begin{figure}
\epsscale{1.0}
\plotone{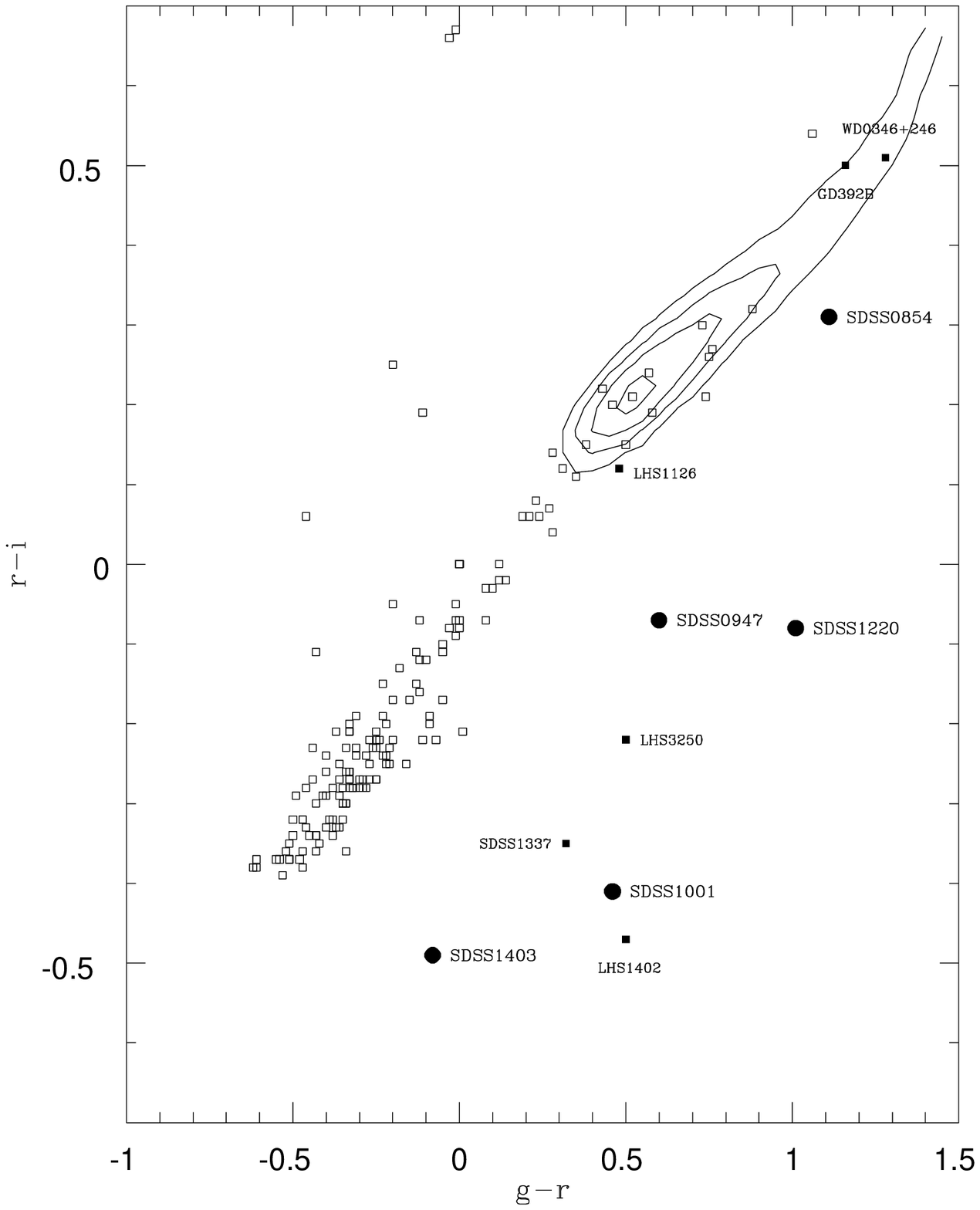}
\caption{Color-color diagram showing five new white dwarfs 
(solid circles) and previously known ultracool white 
dwarfs (solid squares) for which we were able to
estimate SDSS colors (see text for more details).  
A sample of normal white dwarfs (open squares) and contours 
which show the colors
of nondegenerate stars in SDSS are included for comparison.}
\end{figure}


\begin{figure}
\epsscale{0.8}
\plotone{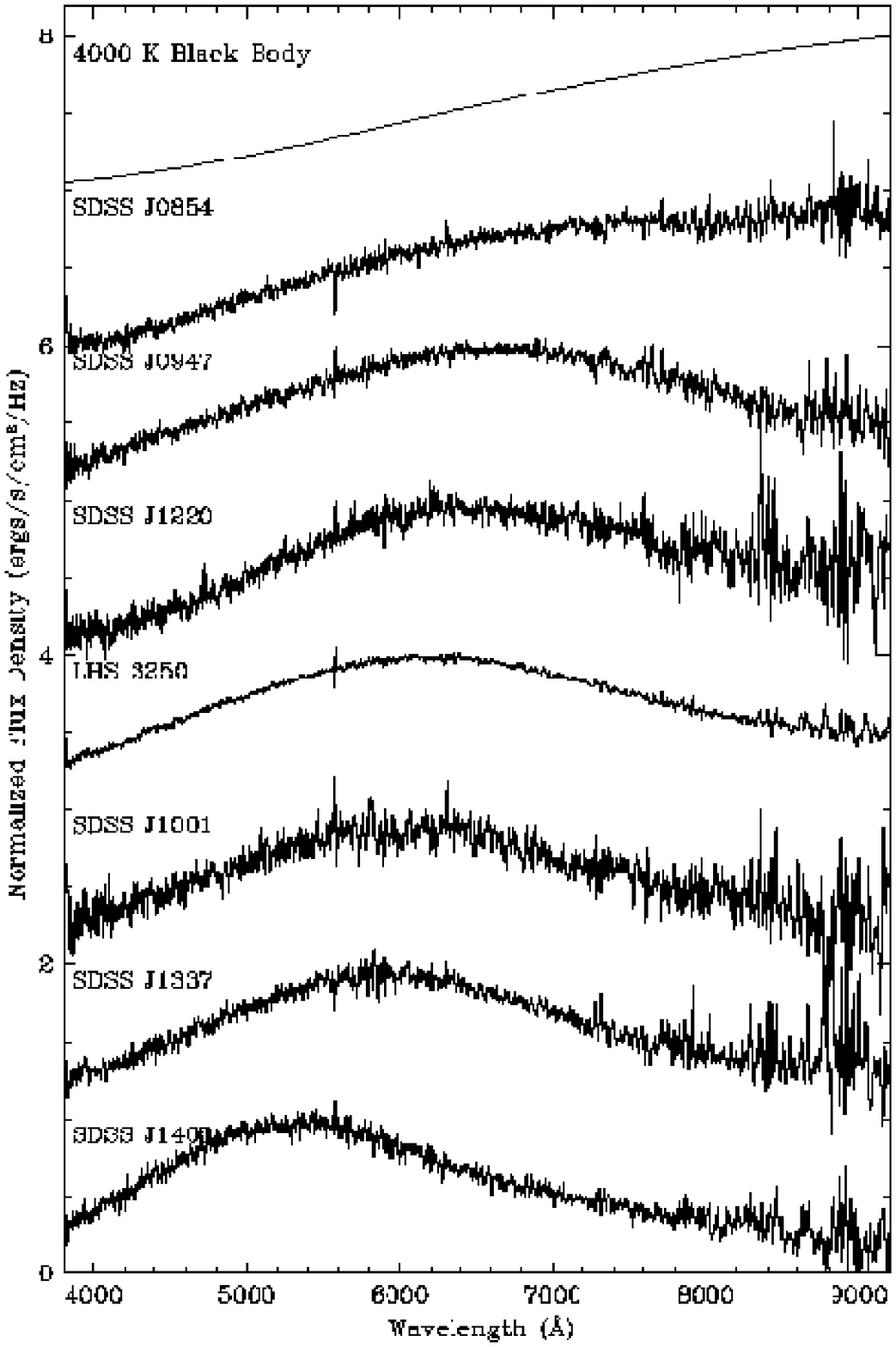}
\caption{Spectra for seven ultracool white dwarf
stars observed in the the SDSS data, including five new stars. Spectra are offset vertically from each other and a 
4000 K blackbody SED is also shown for comparison.
(Spectra have been smoothed by 5 pixels.)  
\label{fig3}}
\end{figure}






\clearpage

\begin{deluxetable}{crrrrr}
\tabletypesize{\scriptsize}
\tablecaption{Observational Data}
\tablewidth{0pt}
\tablehead{
\colhead{Parameter} & \colhead{SDSS J0947} & 
\colhead{SDSS J1220} & 
\colhead{SDSS J1001} & \colhead{SDSS J0854} & \colhead{SDSS J1403}
}
\startdata
RA & 09 47 23.0 & 12 20 48.7 & 10 01 03.4  & 08 54 43.3 & 14 03 24.7\\
dec & 44 59 49 & 09 14 12 & 39 03 40  & 35 03 53 & 45 33 33\\
$\mu$ (mas yr$^{-1}$) & 86 & 504 & 353 & 223 &  284\\
$\mu_{\alpha}$ (mas yr$^{-1}$) & $74 \pm 4$ &$ -341 \pm 15$ &$ -301 \pm 3$
 &$ -133 \pm 5$ & $-271 \pm 3$ \\
$\mu_{\delta}$ (mas yr$^{-1}$) & $45 \pm 3$ & $-372 \pm 15$ & $-185 \pm 3$       &$ -179 \pm 5$ &  $-84 \pm 3$\\
$u$ & 20.71 & 22.40 & 21.39  & 23.63 & 20.14\\
$g$ & 19.45 & 20.35 & 20.04 & 20.49 & 18.93\\
$r$ & 18.85 & 19.34 & 19.58 & 19.38 & 19.02\\
$i$ & 18.92 & 19.42 & 19.99 & 19.07 & 19.51\\
$z$ & 19.40 & 19.89 & 20.51 & 18.92 & 19.82\\
d (pc) & 21-52 & 28-71 & 28-71 & 30-74 & 19-49\\
$v_{tan}$ (km s$^{-1}$) & 8-21 & 68-170 & 47-119 & 31-78 & 26-66\\ 
Julian Epoch & 2002.023 & 2002.192 & 2002.998 & 2002.850 & 2003.176\\ 
SDSS spectra information: & & & & &\\
MJD-plate-fiber & 52672-1202-33 & 52672-1230-58 & 53033-1356-280 & 
52964-1211-395 & 53115-1467-401\\
\enddata



\tablenotetext{a}{Coordinates are given for equinox J2000.0}

\end{deluxetable}

\end{document}